# The use of deep learning in interventional radiotherapy (brachytherapy): a review with a focus on open source and open data

Tobias Fechter[*a,b,c], Ilias Sachpazidis[a,b,c], Dimos Baltas[a,b,c]

[a] *Division of Medical Physics, Department of Radiation Oncology, Medical Center University of Freiburg, Germany*

[b] *Faculty of Medicine, University of Freiburg, Germany*

[c] *German Cancer Consortium (DKTK), Partner Site Freiburg, Germany*

## Abstract

Deep learning advanced to one of the most important technologies in almost all medical fields. Especially in areas, related to medical imaging it plays a big role. However, in interventional radiotherapy (brachytherapy) deep learning is still in an early phase. In this review, first, we investigated and scrutinised the role of deep learning in all processes of interventional radiotherapy and directly related fields. Additionally we summarised the most recent developments. To reproduce results of deep learning algorithms both source code and training data must be available. Therefore, a second focus of this work was on the analysis of the availability of open source, open data and open models. In our analysis, we were able to show that deep learning plays already a major role in some areas of interventional radiotherapy, but is still hardly presented in others. Nevertheless, its impact is increasing with the years, partly self-propelled but also influenced by closely related fields. Open source, data and models are growing in number but are still scarce and unevenly distributed among different research groups. The reluctance in publishing code, data and models limits reproducibility and restricts evaluation to mono-institutional datasets. Summarised, deep learning will change positively the workflow of interventional radiotherapy but there is room for improvement when it comes to reproducible results and standardised evaluation methods.

*Keywords*:

Interventional radiotherapy, brachytherapy, deep learning, open source, neural network

## 1 Introduction

In the treatment and diagnosis of cancer, medical images play an essential role. With the introduction of convolutional neural networks (CNNs) [1], the widespread availability of powerful graphical processing units (GPUs) and the development of easy to use open-source libraries, deep learning (DL) rose to an important tool, assisting clinical decision making in several image driven fields [2]. Also in radiation oncology DL managed to gain foothold [3]. Applications can be found in almost all processes of external beam radiotherapy, ranging from tumour staging, segmentation and registration to treatment planning, treatment delivery and aftercare. One might assume a similar advancement in interventional radiotherapy (IRT). However, the application of DL in IRT is still in an early phase [4,5]. Among other reasons, this can be attributed to the low IRT patient volume. The fact that DL algorithms require a large amount of data and technical knowledge can be an obstacle, especially for small institutions, to conduct DL studies. Consequently, current studies are limited to small mono-institutional datasets [4]. Learning only from site-specific information can lower robustness and general applicability of an algorithm. Additionally, mono-institutional data, not publicly available, makes it impossible for third parties to reproduce results, as a trained model consists of the code and the information from the training data. Beside the datasets, the fine-tuning of the meta-parameters is crucial for a successful DL model. However, the huge amount of meta parameters can't be recorded in a typical journal publication. To increase reproducibility, a solution would be the publication of code and data as open-source and open-data, allowing others to replicate results and compare algorithms on the same datasets. In addition, the entry barrier for developing new algorithms can be lowered with public data [6], which would increase the scientific competition and method diversity. Due to privacy issues, a publication of patient related data is often not possible. In such a case, the publication of the trained models would be an option. The positive impact of open-source and open-data can be seen from the great success of open-source libraries (e.g. Tensorflow [7], PyTorch [8], ModelHub.AI [9]), initiatives like The Cancer Imaging Archive [10] or the increasing number of challenges targeting important medical problems. In this review article, we focus on two objectives: First, we outline and summarise the most recent developments of DL in the area of IRT and directly connected fields. Our second aim was to highlight publications and projects in and close to IRT that publish code, data or models. Through the manuscript, the reader will get a sense of the potential of DL in IRT, the usage of open-source and open-data in the field and its implications for the scientific community.

---

[*] Corresponding author: tobias.fechter@uniklinik-freiburg.de

## 2 Material and Methods

In this section, we first describe how our literature search was conducted, then give a short overview of the reported figures of merit and close the section by describing how the open source policy of research groups was investigated.

As a basis for our review, we conducted a PubMed search in April 2022 for publications in English language using the search terms in Table 1. In addition, we scanned the references of each found paper and looked for other articles citing papers in our search results, to include publications not covered by our search terms and relevant articles from related fields. In a next step, we screened the abstracts of all articles found. If the content of an abstract indicated a fulfilment of our inclusion criteria (see 2.1), the full text article was further inspected.

During the last years, a number of reviews dealing with IRT and artificial intelligence (AI) were published [3–5,11–15]. Whereas, most of them have a wide scope, with only a small section dedicated to deep learning [4,5,11,12,15] or are specific to one organ site [13,14,16]. We discuss the reviews in the course of this work and extend them by the latest publications. Articles already covered by another review will be mentioned only when their content is of peculiar interest or they contribute publicly available source code, data or models.

It should be mentioned, that we do not claim that the list of articles considered for this review is exhaustive. The amount of articles in this field is enormous and we cannot guarantee that all published works found their way in this review. Nevertheless, we can state that this review depicts the current state-of-the-art and lists all relevant deep learning related enhancements in the field.

## 2.1 Inclusion criteria

In order for a manuscript to be included, it had to fulfil at least the criteria regarding the topic and the algorithm. Only for imaging data related papers, the image modality played a role.

**1. Topic:** For an article to be considered it had to be related to IRT and one of the following fields: image segmentation, image registration, dose prediction, treatment planning, outcome prediction or clinical parameter estimation. With this topic selection we cover the whole IRT workflow from the pre-planning phase to treatment. Many techniques of deep learning (especially in image processing) are general, and can be used in IRT without modifications. As an example, it is possible without a big effort to transfer segmentation techniques developed in the scope of radiology or external beam therapy to IRT. Therefore, this article is not limited to publications dedicated to IRT but includes also articles of related fields that could be applied in the IRT workflow, like the before mentioned segmentation or the registration of images.

**2. Algorithm:** Only articles facilitating deep neural networks with multiple hidden layers were considered. Furthermore, we focused on scientific publications and excluded commercial systems.

**3. Image modality:** As IRT is imaging driven, most articles in this review focus on biomedical image processing. Here we focused on the common modalities: ultrasound (U/S), positron emission tomography (PET), magnetic resonance (MR) imaging (MRI), computed tomography (CT), cone beam CT (CBCT)

*Table 1: PubMed search terms for the different IRT processes covered in this review.*

| Process | Search term |
| --- | --- |
| Segmentation | ((brachytherapy) OR ("interventional radiotherapy")) AND ((segmentation) OR (delineation)) AND ((deep learning) OR (neural network) OR (CNN) OR (artificial intelligence) OR (machine learning)) |
| Registration | ((brachytherapy) OR ("interventional radiotherapy")) AND (("image registration") OR ("deformable registration") OR ("rigid registration") OR ("image alignment")) AND ((deep learning) OR (neural network) OR (CNN) OR (artificial intelligence) OR (machine learning)) |
| Catheter reconstruction | ((brachytherapy) OR ("interventional radiotherapy")) AND (("catheter reconstruction") AND ((deep learning) OR (neural network) OR (CNN) OR (artificial intelligence) OR (machine learning)) |
| Dose prediction and treatment planning | ((brachytherapy) OR ("interventional radiotherapy")) AND (("dose estimation") OR ("dose prediction") OR ("treatment planning") OR ("planning") ) AND ((deep learning) OR (neural network) OR (CNN) OR (artificial intelligence) OR (machine learning)) |
| Outcome and clinical parameter prediction | ((brachytherapy) OR ("interventional radiotherapy")) AND ((staging) OR (radiomics) OR ("classification") OR ("gleason") OR ("toxicity") OR ("toxicity prediction") OR ("outcome prediction") OR ("prediction") ) AND ((deep learning) OR (neural network) OR (CNN) OR (artificial intelligence) OR (machine learning)) |

## 2.2 Reported figures of merit

The quality of generated contours is usually measured by the overlap with a corresponding ground truth. Almost all articles dealing with image segmentation report the overlap in terms of the Sørensen-Dice index (DSC) [17]. As the DSC has a lower sensitivity to errors with small volume, some groups report the Hausdorff distance (HD) and the average surface distance (ASD) in addition. For the localisation of needles and catheters, the average displacement of the shaft and the tip error are reported.

The target registration error (TRE) is a common metric in medical image registration. The TRE gives the average distance between corresponding fiducials or anatomical landmarks in the datasets to be registered. In addition, DSC, HD and ASD can be used to assess to registration quality in the region of a contoured volume.

In the case of classification, typically used figures of merit are accuracy, sensitivity, specificity and the area under a receiver operating curve (AUC). For dose prediction algorithms, dosimetric indices are used: $D_x$ gives the dose received by volume x, in relative or absolute volume units.

## 2.3 Open source and open data policy

Another point we were interested in was whether there are institutions or research groups that publish code, data or models constantly with their manuscripts. For this, we strove to identify research groups with a special open-source policy based on the articles in this review. In a first step, we clustered the articles by research group. Articles were assumed to belong to the same group when they have at least three authors in common or at least 50% of the authors of one article are contained by the author list of another article. We used the authors as a surrogate for the affiliations to define research groups because they are easier to process. This has the drawback that the created research groups do not necessarily reflect real institutions but our experience showed that the mapping resembles real institutions very well and is able to depict cross-institutional collaborations. After all manuscripts were clustered, we investigated the amount of published code, models or data in the clusters.

# 3 Segmentation

In this chapter, we focus on tumour and organs at risk (OAR) segmentation on CT, MR, PET and U/S image datasets as well as the digitisation of needles, catheters and applicators. A review covering the segmentation of pelvic cancers (bladder, rectum, cervix, prostate) using deep learning was recently published by Kalantar et al. [18].

## 3.1 CT

The articles identified for CT image segmentation focus on the segmentation of the clinical target volume (CTV) and OARs in the pelvic region. In addition, only few manuscripts could be found addressing the localisation (reconstruction) of catheters and applicators. At the end of this subsection, two publications dealing with image quality are listed.

Liu et al. [19] presented an adversarial network for segmenting the CTV for cervical cancer. Their algorithm achieved a DSC of 0.88. Slightly lower DSC values were reported by Chang et al. [20] investigating the fine-tuning of pre-trained networks for cervical cancer CTV delineation.

DSC values of 0.81, 0.86, 0.86, 0.66 and 0.56 could be reached with a network trained for CTV, bladder, rectum, sigmoid colon, small intestine segmentation in cervical cancer patients [21]. For segmenting the CTV on post implant CTs for tandem and ovoid's IRT a mean DSC of 0.76 was recently reported [22].

Beside the delineation of the tumour region, several papers facilitating deep learning for OAR segmentation on CT could be identified. A DSC of over 0.92 could be achieved by Mohammadi et al. [23] when contouring bladder, rectum and sigmoid of cervical cancer patients. Seven cervical cancer OARs (bladder, bone marrow, left femoral head, right femoral head, rectum, small intestine, spinal cord) were delineated with a deep network by Liu et al. [24] resulting in DSC values of 0.92, 0.85, 0.91, 0.90, 0.79, 0.83 and 0.83, respectively. Another work [25] investigated the performance of two networks in delineating cervix-uterus, vagina, parametrium, bladder, rectum, sigmoid, femoral heads, kidneys, spinal cord and bowel bag. The models could achieve a DSC, averaged over all organs, of 0.88 on an internal test cohort and of 0.79 on an external test cohort. A similar work, contouring prostate, bladder, rectum, left femur, and right femur in the male pelvic region achieved DSC values of: 0.88, 0.97, 0.86, 0.97, 0.97 [26].

A multi task network for joint segmentation and registration was able to reach an ASD of 1.88 mm, 2.41 mm, 2.78 mm, 1.66 mm for prostate, seminal vesicles, rectum and bladder segmentation [27]. A semi-automatic method dedicated to bowel segmentation by Luximon et al. [28] showed an average DSC of 0.90 with the reference contours

Some papers address the segmentation of the prostate alone, either to introduce technical novelties [29,30] or having an outstanding number of patients (more than 1000) in the training cohort [31]. All of them reported DSC values around 0.90. Xu et al. [32] presented a method to delineate bladder, rectum and prostate bed after prostatectomy. A semi-automatic method for prostate delineation on CT and transrectal U/S (TRUS) images was shown by Girum et al. [33].

A dedicated method for segmenting OARs (prostate, bladder, rectum, femoral heads) on CBCT was presented by Fu et al. [34]. They designed a network utilizing CBCT as well as synthetic MRI information for contouring. The group achieved average DSC values above 0.91 for the segmented OARs. Other publications [35–37] used CT and CBCT data to train their models for bladder, rectum and prostate segmentation with DSC values up to 0.87, 0.81 and 0.76.

A U-net based model was presented by Hu et al. [38] to reconstruct tandem and ovoid applicators in CT-based cervix IRT. The presented model achieved on average a DSC of 0.89 and a tip error of 0.80 mm for ten test cases. Jung et al. [39,40] showed in two publications how to segment interstitial needles and tandem and ovoid, Y-tandem as well as cylinder applicators for gynecological cancer IRT. The evaluation yielded DSC values of 0.93 and a HD around 0.70 mm. Similar results (DSC: 0.92, HD < 1 mm) were presented by Zhang et al. [41] and Deufel et al. [42] for tandem and ovoid applicators. Weishaupt et al. [43] proposed a 2D U-net based approach for the segmentation of titanium needles implanted in the prostate, with a mean tip and shaft error of 0.1 mm and 0.13 mm, respectively.

A convolution neural network to reduce the metal artefacts caused by IRT applicators was presented by Huang et al. [44]. Another work segmented bones in the pelvic region to improve beam hardening correction [45].

### 3.2   MR

Compared to CT, MRI offers a higher soft tissue contrast and therefore allows for a more accurate segmentation. Like for the CT a big part of the methods reviewed deals with the segmentation of tumour and OARs in the pelvic region. A few IRT specific publications focus on catheter, implants or IRT specific anatomy delineation.

Cuocolo et al. [46] provide a general overview of the use of machine learning in prostate cancer MR images. A various amount of algorithms exist for the segmentation of the prostatic gland on MR datasets. The main reason for this are three segmentation challenges [47–49], which provide public datasets and a common leader board for evaluation (Table 2). The objectives of the challenges were to segment the prostatic gland [48] as well as peripheral zone and central region of the prostate [47, 49]. The highest reported DSC values were 0.89 for the prostatic gland, 0.79 for the peripheral zone and 0.93 for the central gland [50], respectively. Other public datasets with multiple MRI sequences from different scanners for segmenting the prostatic gland and its substructures are provided by the Initiative for Collaborative Computer Vison Benchmarking [51,52] and the Prostate MR Image Database (https://prostatemrimagedatabase.com/, accessed January 2022). In the challenges, deep convolutional neural networks could achieve the best segmentation results. A good overview can be found in the review by Gillespie et al. [53]. Since the publication of the review, additional methods came up: Tian et al. [54] showed a graph CNN, achieving a DSC of 0.94 on the Promise12 datasets [48]. Also U-net and V-net variants were able to show a good performance on public datasets [55–57]. Liu et al. [58] achieved a DSC of 0.93 on internal data. The segmentation of the prostate on heterogeneous public MRI data was the topic of two other publications [59,60]. The authors reported an average DSC of 0.92 and a symmetric surface distance of 0.77 mm. A U-net version with low computational costs was developed by Comelli et al. [61]. Attention [62] was used by Lu et al. [63] yielding a DSC of 0.93. Another two articles describe deep learning algorithms for segmenting subparts of the prostate [64,65]. Reproducibility, robustness of prostate zone segmentation and how to overcome limited data were the topic of three articles [66–68]. All algorithms for delineating the prostate zones were able to achieve good ground truth overlap (DSC > 0.80) for all zones but the peripheral one. For the peripheral zone the DSC results range between 0.75 and 0.80.

A deep learning model for bladder wall segmentation is given by Hammouda et al [69]. An overview of bladder segmentation on CT and MR images can be gained from the work by Bandyk et al. [70].

Sanders et al. [71] evaluated multiple machine learning models for delineating prostate, seminal vesicles, external urinary sphincter, rectum and bladder on MRI for prostate cancer IRT. In a follow up study, Sanders et al. [72] assessed a network in a prospective clinical trial, showing that the network produces contours of clinical quality. A comparison of two networks for bladder, rectum and femur segmentation can be found in the paper by Savenije et al. [73]. Except for the urinary sphincter with a DSC of 0.70, the investigated networks could achieve very good results with DSC values equal or above 0.80. Zabihollahy et al. [74] reported DSC values of 0.94, 0.88 and 0.80 for bladder, rectum, and sigmoid delineation in the female pelvic with a two-stage CNN approach.

Yoganathan et al. [75] segmented GTV, CTV and OARs (bladder, rectum, sigmoid, small intestine) of cervical cancer patients with an overall average DSC of 0.72.

A comparison of 295 deep learning algorithms to human observers when contouring prostate, external urinary sphincter, seminal vesicles, rectum, and bladder revealed that deep learning based automatic segmentation algorithms can be more consistent than human observers [76].

*Table 2: Publicly available datasets*

| Name | Modality | Structure | URL |
|---|---|---|---|
| Medical Image Segmentation Decathlon [47] | MR | Prostate | medicaldecathlon.com |
| Promise12 [48] | MR | Prostate | promise12.grand-challenge.org |
| NCI-ISBI 2013 [49] | MR | Prostate | doi.org/10.7937/K9/TCIA.2015.zF0vlOPv |
| Collaborative Computer Vison Benchmarking [51,52] | MR | Prostate | i2cvb.github.io |
| Prostate MR Image Database | MR | Prostate | prostatemrimagedatabase.com |
| ProstateX [77] | MR | Prostate | doi.org/10.7937/K9TCIA.2017.MURS5CL |

A 2D U-net for the detection of low dose rate (LDR) IRT seeds was presented by Nosrati et al. [78,79], achieving a maximum difference of 7 mm compared to standard CT-based seed detection.

Beside organ segmentation, also catheter delineation and reconstruction on MR images for prostate [80] and gynecological [81,82] IRT has been tackled by some groups with deep neural networks. A direct comparison is difficult as authors used different figures of merit in their publications. Zaffino et al. [81] reported a DSC of 0.60 and an average distance of 2 mm, Dai et al. [80] showed a catheter tip error of 0.37 mm and a catheter shaft error of 0.93 mm and Shaaer et al. [82] stated a DSC of 0.59 and an average variation of 0.97 mm.

### 3.3 U/S

A challenging task is the segmentation of structures on ultrasound images. Due to the high noise level and the low contrast, an accurate delineation can be very difficult. To improve the image quality of TRUS acquisitions, He et al. [83] presented a generative adversarial network (GAN) to enhance the image resolution. Several groups tried to tackle the problem of delineating the prostatic gland with deep neural networks. The techniques used to create high quality contours comprise residual neural networks [84], attention modules [85], CNNs alone [57,86–88] or in combination with statistical shape models [89–91], learned shape priors [92], a recurrent network for real-time 2D segmentation [93], a regression network with uncertainty estimation [94] or semi-automatic approaches with manual seed points [33,95]. All of them could create contours with high ground truth overlap (DSC >= 0.90). Orlando et al. [96] investigated the effect of TRUS image quality and the number of training images on the performance of neural networks for prostate segmentation. They showed that even with a small number of datasets good results can be achieved and that the image quality of side-fire probes can significantly affect contour accuracy.

The methods so far focused on the segmentation of the prostatic gland as a whole. Segmenting substructures of the prostate on U/S is very challenging due to the low contrast. However, promising results for segmenting the prostate zones were published recently. With a U-net architecture van Sloun et al. [97,98] achieved DSC values above 0.90 for the whole gland and the central zone and a DSC of 0.86 for the peripheral zone.

Lei et al. [99,100] propose a method to segment multiple OARs on TRUS images. Their experiments yielded DSC values of 0.93, 0.75, 0.90 and 0.86 for prostate, bladder, rectum and urethra, respectively. The deep network presented by Behboodi et al. [101] for uterus segmentation showed promising results for the central part of the organ (DSC >0.70) but poor ones for the border regions.

The methods mentioned in the previous paragraphs, were developed to segment structures on U/S images without implanted IRT needles. The needles usually cause severe artifacts in the images, making the detection of the organ boundaries even more difficult. Girum et al. [102] presented a method incorporating learned shape prior knowledge to outline the prostatic gland with inserted needles with a DSC of 0.88.

Deep learning methods have been also used for the identification of IRT needles on U/S images. In the reviewed papers [103–106], the tip error for prostate IRT needles ranged from 0.44 mm to 2.04 mm and the shaft error was between 0.29 mm and 0.74 mm. Liu et al. [107] presented a U-net for catheter segmentation in prostate IRT and analysed factors that might influence the model performance. Additionally, to the segmentation their algorithm provides a confidence metric to support clinicians. A generic method for the localisation of applicators and needles used in prostate and gynecologic IRT, liver and kidney ablation as well as biopsy was presented by Gillies et al. [108]. Their network was able to detect needles and applicators with an average error in tip localisation of 3.50 mm and a DSC of 0.73.

### 3.4 PET/CT

Kostyszyn et al. [109] presented a network to segment the prostatic tumour on PSMA PET images. The network was evaluated on internal and external cohorts and achieved DSCs above 0.81. Matkovic et al. [110] presented a CNN for prostate and tumour delineation on PET/CT with DSCs of 0.93 and 0.80, respectively. For cervical cancer a DSC of 0.84 could be achieved by Chen et al. [111].

## 4 Registration

Modern imaging techniques can provide detailed morphological and functional information about the tumour and OARs. The information improves diagnosis and facilitates a precise, personalised treatment. However, as each imaging modality has its own strengths, an accurate fusion of the information provided by the different modalities is necessary. Due to distinct image characteristics, the time between imaging (change of organ fillings), changes in patient positions and deformations due to external forces (ultrasound probe), the fusion of images, which is also called image registration, can be challenging. Deep learning based algorithms have succeeded also in the field of image registration, convincing with high accuracy and low running time at inference [112,113]. For the fusion of information in IRT, the registration of multimodal image data is of great importance. In the following, an overview of DL based registration of the most common image modalities used in IRT is given.

### 4.1 MR – U/S

One task that is especially important in IRT of prostatic lesions is the registration of MR images, taken before treatment, to TRUS acquisitions acquired during IRT.

Due to the large appearance differences, a huge problem in multimodal image registration is the measurement of the image alignment quality. Haskins et al. [114] optimised a network to learn a similarity metric for MR – TRUS registration. Networks for image based rigid registration were introduced by Guo et al. [115] and Song et al. [116] with an average surface error around 3.60 mm.

Hu et al. [117] designed a network for affine and deformable registration of MR and TRUS datasets. For training, their method needs labelled data, but inference works with plain images. The network achieves an average TRE of 4.2 mm and a DSC of 0.88. A label-driven approach for deformable registration of MR and TRUS prostate images was presented by Zeng et al. [118] . Their method consists of multiple networks. Two for generating TRUS and MR segmentations of the prostate, one for the affine registration matrix and one for the deformation vector field. The mean TRE was 2.53 mm. A similar approach was chosen by Chen et al. [119]. Their method for deformable registration generates vector fields that map the prostate on MR to TRUS images with an average DSC of 0.87. A network that needs only one contour beside the images was presented by Bashkanov et al. [120], showing a TRE of 4.7 mm. Fu et al. [121] incorporated the finite element method into their network and reached a TRE of 1.57 mm.

Ghavami et al. [122] investigated whether the architecture of the segmentation network in a label driven MR-TRUS registration has an impact on the results and found no statistical differences.

### 4.2 PET/CT – U/S

Sultana et al. [123] used a U-net model to generate prostate contours for driving the registration of PET/CT and TRUS images, whereas the registration process itself was not deep learning based. Their method showed a TRE of 1.96 mm.

### 4.3 MR – CBCT

Fu et al. [124] presented a finite element based method for training MR and CBCT registration with a TRE of 2.68 mm. The approach is similar to the one they used for registering MR and TRUS images [121].

### 4.4 Miscellaneous

Lei et al. [125] used a registration network to predict catheter positions for high dose rate (HDR) IRT. First, the case at hand was registered to an atlas, in a second step a regression algorithm predicted the final catheter positions. The evaluation showed that all plans calculated with predicted catheter positions met the common dose constraints. The contour based registration network achieved DSC values of 0.95 for prostate, 0.86 for urethra, 0.93 for bladder and 0.86 for rectum. Guo et al. [126] introduced a network for aligning a 2D TRUS frame with a 3D TRUS volume without hardware tracking. Their rigid registration method showed an average distance error of 2.73 mm. Saeed et al. [127] trained a network to predict prostate motion due to external forces (e.g. ultrasound probe) with an expected error of 0.02 mm.

## 5 Dose Prediction and Treatment Planning

Recently a review about treatment planning optimization in HDR IRT has been published [128]. Only a short chapter in this work deals with machine learning algorithms in the field. That AI methods are still scarce in the field of IRT treatment planning was also confirmed by a recent debate [129]. Nonetheless, the results in the small number of publications were able to show the potential value of AI to IRT.

Shen et al. [130] applied reinforcement learning to adjust organ weights for the HDR treatment planning optimization and showed that their network was able to improve a plan quality score, that considers sparing of OARs, by 10.7% compared to human generated plans. Deep reinforcement learning was also used by Pu et al. [131] to determine the source dwell time for HDR IRT of cervical cancer and could generate higher quality plans than conventional methods. Fan et al. [132] used automated source positions and dwell time estimation to establish a verification tool for QA. Nicolae et al. [133,134] compared the quality of LDR IRT plans for the prostate created by a machine-learning algorithm to the ones created by human observers and did not detect a significant difference. Aleef et al. [135,136] utilised a GAN to LDR prostate IRT treatment planning. Their plans had a similar quality compared to manually generated plans. Jaberi et al. [137] trained models to compensate for intra-fractional organ deformations in gynecological IRT.

Another application of deep learning is to predict the dose calculated in the planning process. Mao et al. [138] developed RapidBrachyDL, a network to predict the dose for Ir-192-based HDR IRT. Compared to Monte Carlo ground truth, the predicted dose deviated for prostate cases by 0.73% for CTV $D_{90}$, 1.1% for rectum $D_{2cc}$, 1.45% for urethra $D_{0.1cc}$ and 1.05% for bladder $D_{2cc}$. Similar, the work by Villa et al. [139] achieved a mean percentage error of -1.19% inside the prostate. As ground truth, they also used Monte Carlo simulations. For cervical cancer HDR IRT PBrDoseSim was introduced by Akhavanallaf et al. [140]. The network showed a mean relative absolute error of 1.16%. Lei et al. [141] used a registration network to transfer dose maps from an atlas to the case at hand. The examined DVH-parameters of their predicted dose maps showed no significant difference to the clinically used dose distribution.

# 6 Outcome prediction and estimation of other clinical parameters

Using deep neural networks for predicting clinical parameters can be attributed to the field of Radiomics [142]. Either in a way to predict clinical parameters directly with a neural network or to use feature-maps of pre-trained networks in combination with other machine learning algorithms. Currently the latter approach is more common as it requires less training data than the first one. Reviews of Radiomics in prostate cancer were recently published [143,144]. As an application related to IRT, the reviews show deep learning methods successfully predicting the Gleason-score. One of the best performing methods was presented by Chaddad et al. [145], who made use of properties extracted from feature maps and were able to predict the Gleason-score with an AUC > 0.80. Apart from Gleason-score prediction many publication deal with tumour grading (e.g. PI-RADS or ISUP) and tissue classification [146–151]. In 2017, Wang et al. [152] showed the potential of deep learning in differentiating benign and malignant prostate cancer tissue. Castillo et al. [153] compared a deep learning [154] and a Radiomics model for prostate cancer classification, interestingly the classical Radiomics approach outperformed the deep learning method on independent test datasets. A multi-stage computer-aided detection and diagnosis model for prostate cancer detection was presented by Saha et al. [155]. The model achieved an AUC of 0.88. Datasets for estimating the clinical significance of prostate lesions are provided by the ProstateX [77] challenge.

Apart from prostate cancer, deep networks were also used to detect vessel invasion in cervical cancer patients. Jiang et al. [156] and Hua et al. [157] reached AUCs of 0.91 and 0.78 on an internal dataset. To determine the myometrial invasion on endometrial cancer MR images, Chen et al. [158] and Dong et al. [159] used deep networks and achieved an accuracy of 0.85 and 0.79, respectively. Wang et al. [160] differentiated malignant and benign ovarian lesions with an accuracy of 0.87. Urushibara et al. [161] showed that a deep network can classify uterine cervical cancer on a radiologist's level. All methods mentioned so far use MRI for their calculations. CT, histology and grade information was used by Dong et al. [162] to predict the lymph node status in operable cervical cancer patients with an AUC above 0.90. Shen et al. [163] facilitated PET/CT image information to predict local recurrence and the occurrence of distant metastasis after chemo-radiotherapy treated cervical cancer patients with an accuracy of 0.89 and 0.87, respectively. The general application of AI in gynecological malignancies was the topic of multiple reviews in the past two years [11,14,16,164].

The dose distribution in the rectum after combined external beam radiotherapy and IRT was used by Zhen et al. [165] for toxicity prediction. Their network achieved an AUC of 0.89, a sensitivity of 0.75 and a specificity of 0.83.

# 7 Open source and open data

While analysing the collected data we noticed that multiple publications made use of publicly available datasets (14%). One should note that all of the six public datasets (Table 2) contain solely MR images and are intended for segmentation and/or classification of prostatic tissue. Around 16% of the manuscripts come with publicly available code and around 5% published the trained models. When viewing at the years of publication a temporal progress can be seen. The number of publications with code or models increased from three in 2019, to seven in 2020 and eight in 2021. An overview of papers with available code or models is given in Table 3. In total 25 manuscripts were published with code, data or models (Table 2 and Table 3). One can see from the tables that IRT specific code or models are scarce [81,135,136], most of the articles deal with topics also relevant for radiology or external beam radiotherapy.

The clustering of the articles by research group to identify groups with a special open-source policy yielded 108 groups in total. What attracts attention when investigating the distribution of reviewed articles per group is that on the one hand, around 77% of the groups are represented by only one manuscript, on the other hand, one single institution composed almost 10% of the articles at hand. An overview of the articles per group is given in Figure 1. Out of the 26 open source or open data contributions, eight came from institutions with only one article in this review. No institution is represented with more than three open source or data articles. The fifth bar in Figure 1 that shows a group with four open source/data publications represents the initiatives around ModelHub.AI [9] and three challenges [47–49]. From the four institutions with more than four articles, only two published open source or data.

# 8 Discussion and Conclusion

The reviewed articles show clearly that deep learning techniques could positively affect the IRT workflow. Most articles concerning the application of DL in IRT were identified in the field of image segmentation, especially in the delineation of organs. As the delineation of organs in IRT is similar to the delineation in related fields like radiology or external beam radiotherapy, the IRT community profited from activities in those fields. A confounding factor that complicates the segmentation of structures in IRT are artefacts caused by implanted catheters or applicators. We could identify only a few algorithms that could segment images with implants [22,102]. For other IRT specific applications like localisation of catheters and applicators, several DL based algorithms could be included in this review. For automated treatment planning it was shown that deep learning algorithms have the potential to generate plans of the same quality as human observers. The big advantage of such algorithms would be that they are able to compute a plan within a few seconds, whereas a manual plan needs several minutes. However, that the inverse plan optimization in IRT based on AI has not been fully explored, might be due to the increasing computational power which enables a very fast multi-criterial optimization (MCO) and searching for the best optimization solution on Pareto space. If the Pareto space can be properly scanned in a fast and consistent way, the produced results will be always more accurate [129].

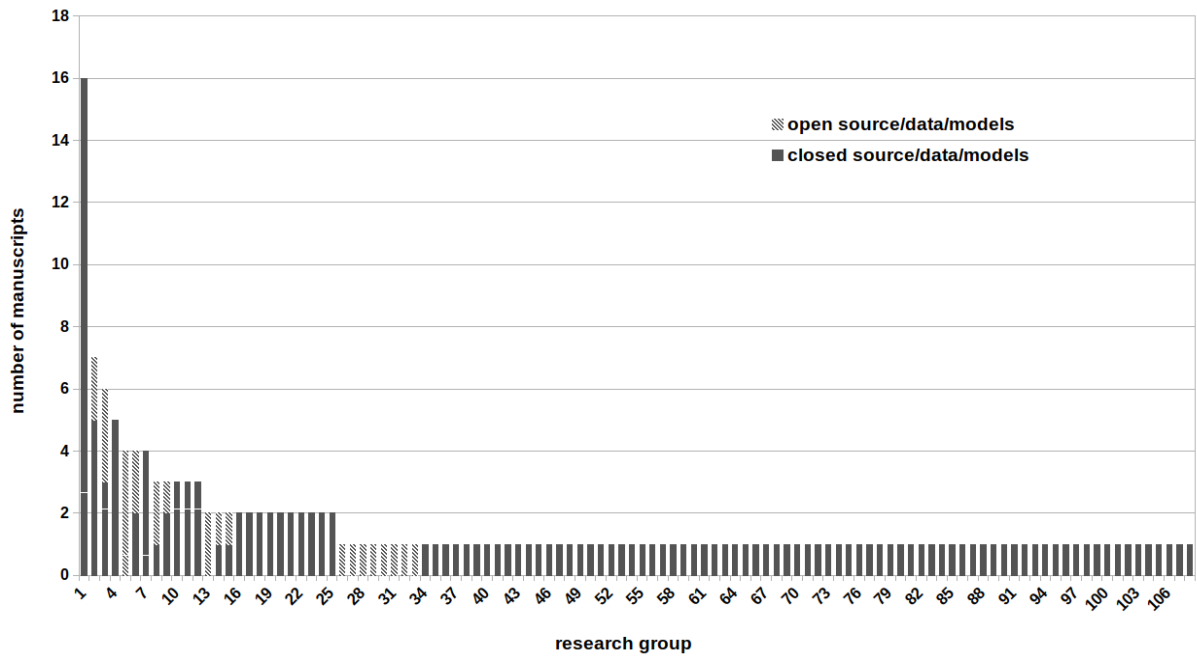

*Figure 1: The height of a bar indicates the number of manuscripts per institution considered in this review. Dark grey represents papers without published code, data or models, bright grey shows articles with either code, data or models.*

The research currently focuses on the pelvic region, which is a main application of IRT and therefore offers more data for the training process. The ratio between articles focusing on the male pelvic area and papers about the female pelvis is approximately 2.5:1. A possible reason for this imbalance are the public challenge datasets for the segmentation or classification of prostatic tissue.

A key factor in the evaluation of deep learning methods is the composition of the cohorts used for training and testing the network. Beside the nominal number of the samples required the quality of the samples play a crucial role. The chosen samples should cover all possible cases. A validation with external independent datasets is important, as it is the only way to prevent an overfitting to institutional specific cohort characteristics [166]. The number of patients in studies was ranged from 10 [130] to 2317 [155]. It was noticeable that most studies applied computational methods, such as k-fold cross validation, to tackle the issue of low patient number and to prevent overfitting. It is difficult to state a number for the minimum amount of samples per training as this is highly task depended. For segmenting the prostate on MR images Saunders et al. [68] showed that, with a proper learning strategy and well selected samples a number of 20 patients can be already enough to generate contours on a human expert level. However, in the same work, the authors showed that changing the learning strategy, can double the amount of data needed to achieve the same contour quality. Most of the reviewed studies limited their research to data coming from a single institution. The restriction to single site data can have severe impact on the generalizability of a model [53,67,68]. All three studies showed that performance can significantly drop when a model is applied to images from a different scanner or patients with other characteristics. Another point that hampers a direct comparison of methods is that many groups reported results only for internal data. The allegedly better performance of an algorithm could be caused by the properties of the used dataset.

*Table 3. List of publications with publicly available source code and models.*

| Name | Scope | Code | Model | URL |
|---|---|---|---|---|
| **Segmentation** | | | | |
| **MR** | | | | |
| Gillespie et al. [53] | prostate | ✓ | ✓ | github.com/AIEMMU/MRI_Prostate |
| Tian et al. [54] | prostate | ✓ | ✓ | github.com/AlanMorningLight/GCN-Based-Interactive-Prostate-Segmentationon-on-MR-Images |
| Liu et al. [59] | prostate | ✓ | | github.com/liuquande/MS-Net |
| Liu et al. [60] | prostate | ✓ | | github.com/liuquande/SAML |
| Savenije et al. [73] | bladder, rectum, femur | ✓ | ✓ | in supplementary of [73] |
| Zaffino et al. [81] | catheter | ✓ | ✓ | available in 3D Slicer [167] DeepInfer [168] |
| Saha et al. [155] | prostate cancer | ✓ | ✓ | github.com/DIAGNijmegen/prostateMR_3D-CAD-csPCa |
| **CT/CBCT** | | | | |
| Xu et al. [32] | bladder, rectum, prostate bed | ✓ | | github.com/superxuang/amta-net |
| Brion et al. [35] | bladder, rectum, prostate | ✓ | | github.com/eliottbrion/unsupervised-domain-adaptation-unet-keras |
| Léger et al. [36] | bladder, rectum, prostate | ✓ | | github.com/eliottbrion/pelvis_segmentation |
| **TRUS** | | | | |
| Wang et al. [85] | prostate | ✓ | | github.com/wulalago/DAF3D |
| Xu et al. [94] | prostate | ✓ | | github.com/DIAL-RPI/PTN |
| **PET** | | | | |
| Kostyszyn et al. [109] | prostate cancer | ✓ | | gitlab.com/dejankostyszyn/prostate-gtv-segmentation |
| **Registration** | | | | |
| **MR ↔ TRUS** | | | | |
| Song et al. [116] | prostate | ✓ | | github.com/DIAL-RPI/Attention-Reg |
| **TRUS ↔ TRUS** | | | | |
| Guo et al. [126] | prostate | ✓ | | github.com/DIAL-RPI/FVR-Net |
| **Planning** | | | | |
| Aleef et al. [135] | LDR plan | ✓ | | github.com/tajwarabraraleef/TP-GAN |
| Aleef et al. [136] | LDR plan | ✓ | | github.com/tajwarabraraleef/3Dpix2pix-for-prostate-brachytherapy |
| **Predicting clinical parameters** | | | | |
| Chaddad et al. [145] | Gleason score | ✓ | ✓ | github.com/fchollet/deep-learning-models |
| Schelb et al. [148] | PI-RADS | ✓ | | github.com/MIC-DKFZ/PROUNET |
| Duran et al. [149] | Gleason score | ✓ | | github.com/AudreyDuran/ProstAttention-Net |

The restriction of many articles to internal data and the fact that deep learning models are only fully reproducible with source code and data available, are the reason why we focused in this work on manuscripts with published code, data or models. The analysis showed that the release of code, data and models is still rare in the field of IRT but increasing over time. The clustering of publications into research groups could not identify any group that publishes code, data or models on a regular basis. However, the clustering revealed that people from one institution authored a big part of the deep learning publications. To the best of our knowledge, neither code, data nor models were released with these manuscripts, which limits validation and reproduction of their results significantly. The dominance of one institution bares the risk that developed algorithms do not reflect the whole spectrum of methodologies. Especially in a field like IRT that relies heavily on human experience, which causes variable techniques across institutions this should be seen critically.

In our eyes, public datasets are the foundation to enable a fair comparison of methods and to keep diversity in the scientific field. In addition to the datasets a common assessment framework with a predefined data processing pipeline is needed (e.g. the significance of a high DSC value is a different one whether only a centre cropped region of interest is handled by an algorithm or a full body scan is taken as input). Beside a common dataset and assessment framework, it is desirable that author publish their code. We see two big advantages in public code. First, it enhances reproducibility. Second, it would allow others to build upon an existing code-base, which can accelerate scientific progress. The European Union has also identified these needs. The EU-funded project ProCancer-I strives to build an open-source framework for the development, sharing and deployment of AI models for prostate cancer patients (https://www.procancer-i.eu/, accessed February 2022). Other noteworthy projects that pursue the same objectives are e.g. The Cancer Imaging Archive [10], Image Biomarker Standardization Initiative [169], ModelHub.AI [9], MONAI [170], etc.

A limitation of our current review might be that it has a big focus on image processing. However, we think that the review shows clearly the big potential of deep learning in IRT and the role of open source, models or data in the field. Summarised, we can state that the current scientific progress will change positively the workflow of IRT and there is room for improvement when it comes to reproducible results and standardised evaluation frameworks.